%% file: _main.tex
\documentclass[sigconf]{acmart}
\usepackage{booktabs} % For formal tables

\setcopyright{rightsretained}
\copyrightyear{2018}
\acmConference[preprint]{preprint}{v.2, November 2020}{}

\begin{document}
\title{Between an Arena and a Sports Bar: Online Chats of Esports Spectators}
% \subtitle{Extended Abstract}

\author{Denis Bulygin}
\affiliation{%
  \institution{National Research University Higher School of Economics} 
  \city{St.Petersburg, Russia}
}

\author{Ilya Musabirov}
\affiliation{%
  \institution{National Research University Higher School of Economics}
  \city{St.Petersburg, Russia}
}

\author{Alena Suvorova}
\affiliation{%
  \institution{National Research University Higher School of Economics} 
  \city{St.Petersburg, Russia}
}

\author{Ksenia Konstantinova}
\affiliation{%
  \institution{National Research University Higher School of Economics}
  \city{St.Petersburg, Russia}
}

\author{Pavel Okopnyi}
\affiliation{%
  \institution{University of Bergen}
  \city{Bergen, Norway}
}

\renewcommand{\shortauthors}{D. Bulygin et al.}

\begin{abstract}
Hundreds of thousands of spectators use Twitch.tv to watch The International, a Dota 2 eSports tournament and communicate in massive chats.
In this paper, we analyse crowd behavior in these chats, disentangle features of social communication, such as contextual meanings of emojis and short messages. We apply structural topic modelling and cross-correlation analysis to investigate topical and temporal patterns of chat messages and their relation to in-game events. We show that in-game events drive the communication in the massive chat and define its emergent topical structure to a various extent. Following the discussion in communication and social computing literature, we discuss these findings in the framework of analysis of communication of physical sports crowds and outline some limitations of the 'stadium' metaphor, suggesting a complementary metaphor of 'sports bar' as a useful analytical and design device.
\end{abstract}

\begin{CCSXML}
<ccs2012>
 <concept>
  <concept_id>10010520.10010553.10010562</concept_id>
  <concept_desc>Computer systems organization~Embedded systems</concept_desc>
  <concept_significance>500</concept_significance>
 </concept>
 <concept>
  <concept_id>10010520.10010575.10010755</concept_id>
  <concept_desc>Computer systems organization~Redundancy</concept_desc>
  <concept_significance>300</concept_significance>
 </concept>
 <concept>
  <concept_id>10010520.10010553.10010554</concept_id>
  <concept_desc>Computer systems organization~Robotics</concept_desc>
  <concept_significance>100</concept_significance>
 </concept>
 <concept>
  <concept_id>10003033.10003083.10003095</concept_id>
  <concept_desc>Networks~Network reliability</concept_desc>
  <concept_significance>100</concept_significance>
 </concept>
</ccs2012>
\end{CCSXML}

\ccsdesc[500]{Human-centered computing~Empirical studies in collaborative and social computing}
\ccsdesc[300]{Information systems~Chat}

\keywords{Streaming, eSports, chat communication, Twitch.tv}

\maketitle

\input{_text}

\bibliographystyle{ACM-Reference-Format}
\bibliography{2018chi}

\end{document}

%% file: _text.tex
\section{Introduction}

Millions of people watch sports events in stadiums, arenas, sports bars, and at home. In the esports world, in which the sport is happening in a virtual space, public viewing places are replaced with streaming platforms. In these platforms, thousands and even millions of viewers watch an event and share their opinions and experiences via text chat. 

In this work, we study communication between viewers of The International (TI), one of the most significant esports events in the world, and the biggest annual tournament in Dota 2, a team-based multiplayer online battle arena. 

We used chat data from the main broadcast channel on Twitch.tv during TI7 (2017) to study what is called ``crowdspeak''\cite{ford_chat_2017} -- a meaningful and coherent communication of thousands of viewers in the chat.

First, we investigate the thematic structure of viewers communications and disentangle contextual meanings of emotes and text shortcuts using Structural Topic Modelling (STM)\cite{roberts_stm_2019}.

Second, we explore the connection between game events and topics occurring in the chat using cross-correlation analysis. In-game events to some extent define the topical structure of the chat -- they provoke emotional response or discussion of players and teams, while the lack of action on screen leads to viewers frustration which is expressed with boredom-related memes and emotes. 

Last, we unveil the nature of the inequality between topics in the chat. In larger chats, participants tend to focus on fewer topics while in smaller chats a variety of discussion points can be found. As the tournament progresses, viewers become more emotionally engaged and focused on cheering, omitting other topics. 

Based on our findings, we propose design ideas aimed to enhance viewers' experience.

\section{Related Works}

Groups of ``like-minded fans'' watch together sports events in public spaces, such as a pub or a bar, and cheer for their favourite teams and athletes ritually \cite{weed_pub_2007}. As fans share their experience with each other in various forms, spectatorship becomes an inherently social activity \cite{rubenking_sweet_2016}. Reeves et al. \cite{reeves_designing_2010} describe the transformation of crowd behavior with increase of crowd size in real-life settings, discussing what constitutes the crowd, and outlining design challenges for this task, known as design for crowds. They highlight that crowd consists of an active minority of a lot of cliques which synchronize through having common objects: dress elements, songs, body movements, etc. 

These findings are also connected to evolutionary behavioral science research on different kinds of synchronous behavior as ways to overcome the limits of human cognitive ability to sustain social connections\cite{dunbar_social_1995,dunbar_online_2016}. Cooperative activities, such as dancing, are shown to help to overcome these limitations and increase pro-social attitudes \cite{pearce_singing_2016,reddish_collective_2016,atherton_walking_2020}.

Flores-Saviaga et al. \cite{flores-saviaga_audience_2019} uncover five types of streams with regard to their audience size and  activity. Types of audience activity is related to a number of active participants: the more people are active in the chat, the less messages each particular user writes and the shorter they are. 

Viewers provide each other with information cues on how to behave oneself in a chat forming a ``normative behaviour'' \cite{cialdini_social_2004,vraga_filmed_2014}, patterns of which in massive chats are quite different from ones in smaller chats \cite{ford_chat_2017}
. Messages flow rapidly, forming a ``waterfall of text'' \cite{hamilton_streaming_2014}, which makes it almost impossible to read messages one-by-one and have a meaningful conversation. through seeking of cohesion with using common emotes and copypastas.

When an interesting event happens in a game (e.g., a death of a hero in our case), viewers react ``loudly'': they post a burst of messages, creating a ``local meaning of the [in-]game event'' \cite{recktenwald_toward_2017}. Viewers copy and paste, or type emotes, abbreviations, and memes, launching cascades of messages which disrupt the usual flow of communication \cite{seering_shaping_2017} causing a ``communication breakdown'' \cite{recktenwald_toward_2017}.

Communication in massive chats is far from meaningless. Users engage in various practices which ensure chat coherence. They post and re-post abbreviations, acronyms, and emotes in chat, creating a ``crowdspeak'' in which messages from many viewers unite into ``voices'' \cite{ford_chat_2017, trausan-matu_polyphonic_2010} -- particular positions or discussion threads adopted and expressed by many participants. The most straightforward approach in the operationalisation of voice is to consider a repetition of a word or a phrase to be a voice and the number of repetition to be its strength\cite{trausan-matu_polyphonic_2010}. Ford et al. \cite{ford_chat_2017} hypothesised that massive chats would have fewer unique voices in comparison to smaller chats; however, this hypothesis was not confirmed nor rejected.

\section{Background}

Dota 2 is an online multiplayer game in which two teams of five players compete for domination over the game field (map). Confrontation involves elimination of opponent team's characters (heroes) which later respawn to continue the fight. A typical game lasts for approximately 15 to 45 minutes\footnote{https://dota.rgp.io/\#chart-avg-duration}. During esports events, teams confront each other in matches, which consist of 1 to 5 games each.

TI7, like most sports tournaments, has several stages: qualifiers, groups, playoff, and finals (see Table \ref{tab:stages} for details). Qualifiers decide which teams get accepted to the tournament. During group stages, the initial position of the team for the playoff is decided. In the playoff, losing teams are eliminated from the tournament. In the end, two teams compete in the finals.

\section{Data and Methods}

We employed the Chatty \footnote{chatty.github.io/} application to record chat message of the \textbf{dota2ti} channel on Twitch.tv which broadcasted most of the matches of TI7. In total, we collected more than 3 million chat messages from approximately 180 thousand unique viewers. We complemented these data with information on in-game events, which we obtained by employing Open Dota 2 API and the Dotabuff.com database.

\begin{table}[h]
\small
\caption{TI7 stages}
\begin{tabular}{lrrr}
\textbf{Stage}                      & \textbf{Groups} & \textbf{Playoff} & \textbf{Finals} \\
\toprule

\textbf{Days}                       & 4               & 5                & 1               \\
\textbf{Games}                       & 100               & 43                & 7               \\
\textbf{Messages}                   & 819857          & 1831529          & 381690          \\
\textbf{Mean msg. length (SD)}      & 30 (47)         & 26 (44)          & 32 (47)         \\
\textbf{Documents}                  & 29378           & 36165            & 5672            \\
\textbf{Mean doc. length (SD)}      & 867 (726)       & 1405 (962)       & 2274 (1387)     \\
\textbf{Viewers}                    & 78106           & 128278           & 59070           \\
\textbf{Mean per viewer}            & 10 (38)         & 14 (56)          & 6 (16)          \\
\textbf{Share of emotes}            & 32\%            & 36\%             & 28\%            \\
\textbf{Share of mentions}          & $\sim$1\%       & $\sim$1\%        & \textless 1\%   \\
\textbf{Mean viewers per game (SD)} & 2573 (3444)     & 8084 (15225)     & 14490 (14919)   \\
\textbf{Mean msgs. per game}         & 8199            & 42594            & 54527  \\
\bottomrule
\end{tabular}
\label{tab:stages}
\end{table}

\subsection{Structural Topic Modelling}

We applied Structural Topic Modelling (STM) \cite{roberts_stm_2019,roberts_structural_2013} to analyze the contents of the chat. STM shares the same basic approach with other probabilistic topic models (e.g. Latent Dirichlet Allocation \cite{blei_probabilistic_2012}): it takes a corpus of text documents as an input and produces a given number of topics -- groups of words which occur in text often together.  

STM, however, can take into account connection of topics probability with document-level metadata, which in our case was the stage of the tournament the game belongs to: Groups, Playoff, and Finals.

Chat messages on Twitch.tv are very short, often consisting of one or several words, emotes or shortcuts, which is not suitable for probabilistic topic modelling. We concatenated messages into documents, each covering a 7 second time window (mean = 7.99 messages per second, SD = 7, max = 115). 

\subsection{Analysis of Event-driven Nature of Communication}

We applied cross-correlation analysis \cite{brockwell_time_2013} to investigate the connection between in-game events and topics prevailing in the chat. In this work, for a given time window, we consider a topic to prevail in case it is the most frequently occurring according to the STM results. For each time window, we also calculated the number of happened in-game events: usually, 1 or 0. We tested the resulting time-series with the Kwiatkowski-Phillips-Schmidt-Shin (KPSS) test to ensure they are stationary \cite{kwiatkowski_testing_1992}.

Cross-correlation analysis tests if there is a correlation between two time series with a lag in some range. It produces a vector of correlation coefficients, showing whether there is a tendency for events in one time series to precede, follow, or occur concurrently with the events in another.

As a result, for each of 110 topics, we computed a vector of correlation coefficients between two time series - in-game events and topic-prevalence - with lags in seven-seconds time frames. These temporal patterns show us if topics in the chat are preceded, followed, or prevail in the chat during the in-game event.

Having 110 vectors of correlation coefficients, we united them into groups of similar patterns. We used Spearman's correlation as a measure of similarity between vectors and applied hierarchical clustering to produce the groups.

\subsection{Analysis of Voices and Topical Inequality}

To analyse the connection between the context (tournament stage), the number of participants and unique voices\cite{trausan-matu_polyphonic_2010} in chats, we propose an alternative to Ford et al. operationalisation \cite{ford_chat_2017}, treating STM topics as proxies to unique voices and looking at the topical inequality\cite{zeileis_ineq_2014}. 

As a result of STM we have topic distribution for each 7-second time window. So the document can be homogeneous (there is one prevalent topic in this time window), can have several important topics or can be extremely heterogeneous (many equally important topics with low frequencies). To summarize the distribution of topics on the game level we calculate cumulative topic distribution and normalize it to account for different number of 7-second periods per game. 

For every game, we calculated Gini coefficient for topic distribution in the chat during the game. Gini coefficient ranges from 0 to 1, where 0 means absolute equality among topics (all topics are represented equally in the chat) and 1 means that only one topic is present while others are missing. We treat Gini coefficient as an estimate of the extent to which some voices prevail (are ``stronger''\cite{trausan-matu_polyphonic_2010}) in the chat over others.

Our analysis of this inequality is two-fold: (1) expanding Ford et al. approach to the whole text corpora, we look at Gini coefficient distribution in connection with the number of the particular game spectators, (2) we test for significant changes in Gini coefficients  between different tournament stages.

Ford et al. suggested that massive chats would contain fewer unique voices and be less polyphonic. In our case, the number of voices (i.e., the number of topics) is predefined and constant. However, we can estimate and test the relation between the number of viewers in chat and topical inequality. 

We applied Spearman's correlation test to test the hypothesis that topics in larger chats would be less equally distributed. 

We also assessed differences in Gini coefficient between stages using Kruskal-Wallis test and post hoc Dunn's test to investigate if the topical inequality depends on the context of the chat.

\section{Analysis and Results}

\subsection{Event-driven Nature of Communication}

After clustering topics according to their temporal patterns, we resulted in four groups of topics which we labeled based on their contents: 
\textbf{Professional Scene} (1 cluster),
\textbf{Boredom and Frustration} (2 cluster), \textbf{Emotional Response} (3 cluster), and \textbf{Game and Stream Content} (4 cluster). 

\begin{figure}
  \centering
  \includegraphics[width=0.46\textwidth]{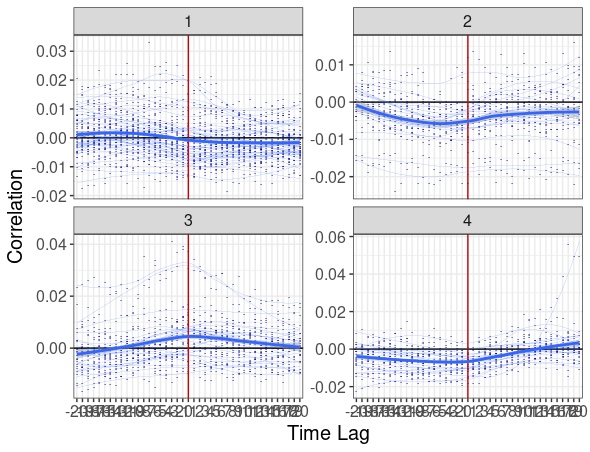}
  \caption{Temporal patterns of groups of topics}
  \label{fig:clusters}
\end{figure}

Association between these clusters and the topic model captures the differences in a context for similar topics and particular tokens (words, emotes), which is an especially fortunate feature for an emote-based contextually rich communication of streaming chats.

\paragraph{Boredom and Frustration.}

When nothing happens in the game, viewers are bored and convey this boredom and frustration in various ways. They send specific emotes  (e.g. \textit{ResidentSleeper}) or even start ``copypasta'' cascades by copy-pasting certain emotes and messages.

When no in-game events are happening, topics of this cluster are in the chat. When the tension starts building and viewers anticipate interesting events to happen, viewers stop sending boredom-related messages. During and after the event, the level of these topics in the chat remains low.

\paragraph{Emotional Response.}

This cluster of topics represents spectators' response to in-game events: the death of a character, destruction of a building, or a cameraman missing an in-game event, for example. Messages often contain only one word or emote. Viewers write ``gg'' (abbreviation of ``good game''), ``ez'' (short handing for ``easy'') at the end of a match, ``322'' (a Dota 2 meme\footnote{http://knowyourmeme.com/memes/322}) to mock a player or team due to their poor performance, emotes ``\textit{pogchamp}'' (glory) and ``\textit{kreygazm}'' (excitement) to convey their feelings to what is happening on in the game.

These messages appear and fill the chat soon before an in-game event and disappear shortly after, as viewers anticipate something happening and then react loudly.

\paragraph{Game and Stream Content.}

Topics in this cluster reflect viewers' reaction to whatever is happening on screen at the moment. They discuss and cheer teams and players. Even though players can not perceive the audience, viewers still address their messages to them: ``BudStar BudStar BudStar LETS GO LIQUID BudStar BudStar BudStar''.

Topics of this cluster appear in the chat mostly after in-game events. In general, spectators comment and discuss stream content during the whole stream except for moments when in-game situations grab their attention.

\paragraph{Professional Scene.}

Topics of this cluster do not significantly relate to in-game events. Viewers cheer professional players and teams, discuss their in-game behaviour and even past incidents involving players almost all the time, and we did not notice any temporal pattern associated with these topics. 

Besides listed, there are, of course, other topics in the chat which are less loud. They are included in the aforementioned four groups. Thus, our interpretation of each group is based on the prevailing content of its topics and does not necessarily take into account all of the variety of themes and discussion objects that can be found in the chat.

\subsection{Voices and Topical Inequality}

\begin{figure}
  \centering
  \includegraphics[width=0.46\textwidth]{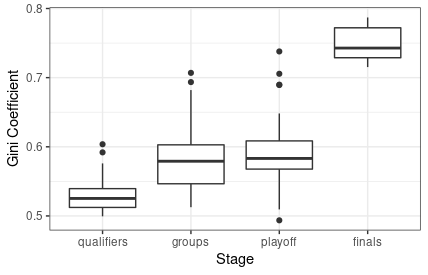}
  \caption{Gini coefficient during different stages of TI7}
  \label{fig:giniInitial}
\end{figure}

The finals of any sports tournament attract lots of people who come to cheer their favourite team, and TI7 was not an exception. Kruskal-Wallis test showed that there is a significant difference ($H = 58.9, df = 3, p  < 0.001$) in Gini coefficient among stages. Using Dunn's test with Bonferroni correction for post hoc analysis, we found that the stages differ significantly except the comparison between Playoff and Groups stages (See Table \ref{tab:topic53}, part a) and Gini coefficient increases with the tournament's progress (See Fig. \ref{fig:giniInitial}).

\begin{figure}
  \centering
  \includegraphics[width=0.46\textwidth]{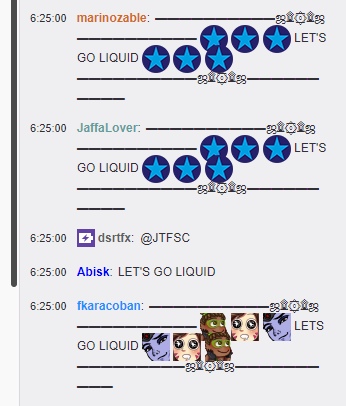}
  \caption{Cheering for Team Liquid during Finals}
  \label{fig:letsgo_liquid}
\end{figure}

Further exploration showed that 70\% of messages during Finals belong to the single topic (topic 24) which is dedicated to cheering for one of the finals participants (See Fig.\ref{fig:letsgo_liquid}). Closer to the final game of the tournament cheering increases, displacing all other voices and discussion threads.

\begin{figure}
  \centering
  \includegraphics[width=0.46\textwidth]{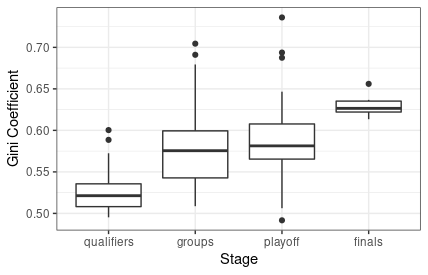}
  \caption{Gini coefficient during different stages of TI7}
  \label{fig:topic24}
\end{figure}

When we removed topic 24 from calculations, we could no longer claim significant differences between Finals and Playoff, while the other differences remain significant (Finals and Groups, Qualifiers and all other stages -- see Fig. \ref{fig:topic24} and Table \ref{tab:topic53}, part b, Kruskal-Wallis test: $H = 57.1, df = 3, p  < 0.001$).

\begin{table}[h]
\caption{Pairwise comparison of Gini coefficient of games on different stages (Dunn's test)}
\small
\begin{tabular}{lcc}
 & \textbf{(a) Topic 24 Included} & \textbf{(b) Topic 24 Excluded} \\
\textbf{Finals - Playoff} &	 $z = 2.68, p = 0.021$	 & 	$z = 2.06, p = 0.117$	 \\
\textbf{Finals - Groups} &	 $z = 3.26, p = 0.003$	 & 	$z = 2.72, p = 0.019$	 \\
\textbf{Finals - Qualifiers} &  $z = 5.77, p  < 0.001$	 &  $z = 5.27, p  < 0.001$	 \\
\textbf{Playoff - Qualifiers} &  $z = 6.15, p  < 0.001$	 &  $z = 6.37, p  < 0.001$	 \\
\textbf{Playoff - Groups}  &  $z = 1.16, p  = 0.733$	 &	$z = 1.32, p  = 0.559$	 \\
\textbf{Groups- Qualifiers} &  $z = 4.84, p  < 0.001$	 &  $z = 4.91, p  < 0.001$	 \\
\end{tabular}
\label{tab:topic53}
\end{table}

Thus, during the Finals, a single topic was dominating the chat, reducing unique voices to variations of chanting for the favourite team.

\begin{figure}
  \centering
  \includegraphics[width=0.46\textwidth]{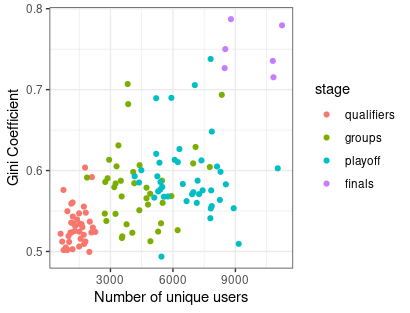}
  \caption{Gini coefficient and number of participants}
  \label{fig:sizeGini}
\end{figure}

There is a clear tendency that later stages of the tournament attract lager audience (more users watch finals comparing to number of users watching games during groups). At the same time we found that Gini coefficient increases with the tournament's progress and chat became less polyphonic (Fig. \ref{fig:sizeGini}). So, to explore the relationship between topic inequality and chat size we need analyse stages separately.

\begin{figure}
  \centering
  \includegraphics[width=0.46\textwidth]{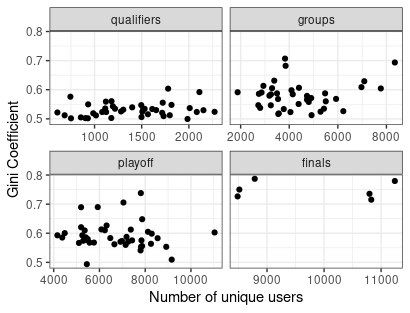}
  \caption{Gini coefficient and number of participants per stage}
  \label{fig:groups_corr}
\end{figure}

 For each stage we perform Spearman's correlation test for number of unique participants in each game and Gini coefficient for topics in that game. The test showed that there were no significant correlation between the size of the chat and the Gini coefficient (see Fig. \ref{fig:groups_corr} and Table \ref{tab:corr}). Thus, topics distributions do not differ between larger or smaller chats during the same stage.

\begin{table}[h]
\caption{Correlation between number of unique chat participants and Gini coefficient}
\small
\begin{tabular}{lcc}
\textbf{Stage} & \textbf{$\rho$} & \textbf{p-value} \\
Finals & 0.086	 	 & 	0.919	 \\
Playoff & -0.198	 	 & 	0.207	 \\
Groups & 0.013	 	 & 	0.940	 \\
Qualifiers & 0.228 	 &  0.151	 \\
\end{tabular}
\label{tab:corr}
\end{table}

\subsection{Prevalent Topics Over Stages}

To compare stages of TI7 with each other, we compiled a list of prevalent topics which are associated (based on STM model) with the given stage.

\begin{figure}
  \centering
  \includegraphics[width=0.46\textwidth]{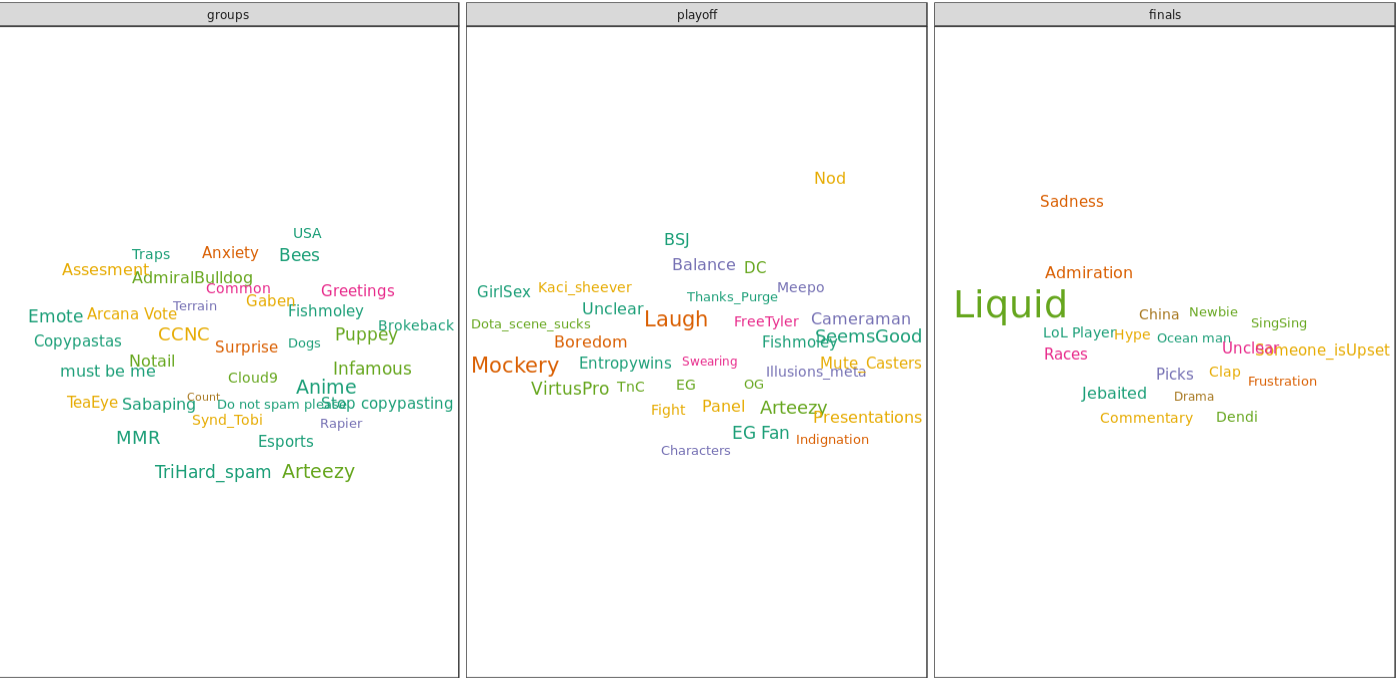}
  \caption{Topics prevalent in TI7 stages. Size represents the share of messages on the topic}
  \label{fig:topic_shares}
\end{figure}

Group stage has the set of 34 prevalent topics (see Fig. \ref{fig:topic_shares}). Viewers mostly discuss broadcast-related issues (yellow) and famous players (green). They copy-paste texts unrelated to the stream content (e.g.  No job \textit{4Head} No girlfriend \textit{4Head} No friends \textit{4Head} No talents \textit{4Head} Wasting time on Twitch \textit{4Head} Must be me ) (cyan). The expression of emotions is present (brown) but does not stand out from other topics.

During Playoff (31 prevalent topics), in which teams are getting eliminated, chat communication becomes more focused on the game. Viewers discuss game elements: balance or position of the camera (violet), and actively express emotions (brown).

In the Finals (19 prevalent topics), more than a half of all messages were expressing support to Team Liquid -- a western team which was opposing team Newbie from China. The emotional response to events is also present (brown) while forms of copypasta other than chanting almost disappear.

\section{Discussion}

In-game events, the context of the game -- all of these factors contribute to the topical structure and contents of the chat during the stream, offsetting the effects of  number of participants in the chat. 

Using Topic Modelling and statistical analysis methods we reveal some important factors behind the viewers' behaviour in the chat. We show that the chat is overall event-driven: in-game events or the lack of those define the contents of strong voices heard in the chat. The context of the game affects the strength of heard voices: the closer to Finals, the stronger gets the cheering while other voices fade.

The crowd adapts participation practices to characteristics of communication flow and context, which is driven by game events. E.g. copypasta on the earlier tournament stages can convey boredom and frustration, but when the tournament tension goes up, it is used mostly to cheer the favourite team. Thus, the chat becomes a tribune, an arena, or a sports bar, in which visitors watch the game and engage in discussions or cheer for their favourite team loudly. This corresponds to many characteristics of real-life sport crowd behavior\cite{reeves_designing_2010}, while the instruments and practices of synchronisation change to virtual ones.

In this context, we believe the practices of coherence seeking through copypasting behavior described by Ford et al.\cite{ford_chat_2017}, and confirmed to have various event-driven nature in this work, can be a way for a crowd to synchronize in the situation of shortage of physical common objects, role of which is explored in \cite{reeves_designing_2010}.

These findings correspond to the similar position of evolutionary behavior research on practices of synchronization and their pro-social influence. Dunbar et al. \cite{dunbar_social_1995,dunbar_online_2016} reveal that the limits of cognitive ability of humans to meaningfully contribute to the conversation and strengthen ties in the group both offline \cite{dunbar_social_1995} and online \cite{dunbar_online_2016} peak at 4-5 participants in the group (cliques). Nevertheless, human can overcome these limitations and increase close relations in larger groups by doing cooperative activities, leading to synchrony, such as dancing \cite{pearce_singing_2016,reddish_collective_2016}. These activities are shown to be able to increase pro-social behavior, both to other experience participants, and out-groups \cite{atherton_walking_2020}.

In this way, different types of reactions to events we describe, appear to be an example of synchronization natural to human behavior, with the context of the streams framing what practices are more suitable and natural to participants: there are examples of practices for reacting to the events on screen (particular emotes and short words) or lack thereof (copypastas and emotes expressing boredom). 

\section{Design Implications}

The event-driven and sentiment sharing nature of massive tournaments' chats suggests rethinking its design goals.

Reeves et al. \cite{reeves_designing_2010} in their analysis of football spectators crowds also make relevant, in our opinion, call for switching the lens of participation from the level of crowd as composition of individuals to treatment of crowd as \emph{crowd per se}, in other words, an emergent social phenomenon, produced by non-linear aggregation of micro-behavior of individuals. Further analysis of this emergent phenomenon will allow to focus on social coordination mechanisms, leading to the observed behavior. In distinguishing virtual crowds from physical, design can borrow from the exploration of the role of common text patterns (copypastas, emojis) as virtual analogs of shared objects, which Reeves et al. \cite{reeves_designing_2010} treat as playing important role in synchronization.  

While we suggest that the experience of spectators is consistent with a sports arena metaphor~\cite{hamilton_streaming_2014,ford_chat_2017}, it appears to contain characteristics of various types of experiences, outlined by \cite{reeves_designing_2010} for football spectating.

In relation to sports arena metaphor, massive Twitch chat experiences appear to be mediated by the same intra-audience effect \cite{cummins_mediated_2017} that emerges during live spectating, and plays its role  in enhancing co-spectators perceptions. 

In addition, we also observe that some traits of a sports bar metaphor are also accurate since players usually do not receive direct feedback from chat participants during the game. While shouts and chants do not reach the addressee, spectators still find these responses important \cite{ford_chat_2017}, and thus the metaphor can lead to complementary design ideas. 

Based on our results, we would like to propose two design implications to enhance the chat participants experience. 

\subsection{Highlights}

Due to its event-driven nature, the chat can be a valuable and reliable source of information regarding in-game events: which are most notable, intense, or funny. A record of the game can be automatically cut (see \cite{wang_automatic_2005} for examples) into a short, meaningful clip which would convey all vital information about the game for those viewers who want to re-experience it.

\subsection{Visualization of Prevailing Voice}

As tension grows during the tournament and the topical inequality rises in the chat, certain topics become dominant and occupy significant screen space, leaving almost no place for other topics/voices. During the Finals of TI7, this topic was dedicated to cheering for the Team Liquid (see Fig. \ref{fig:letsgo_liquid}). 

We suggest providing additional instruments of sentiment-sharing in the form of graphical elements or counters which would indicate the current sentiment of the chat. These indicators will inform users of the loudest voices in the chat, provoking to join one of them and participate in the coherent practice. 

\begin{acks}

The article was prepared within the framework of the Academic Fund Program at HSE University in 2020 — 2021 (grant No. 20-04-024) and within the framework of the Russian Academic Excellence Project «5-100».

\end{acks}